\documentclass[jkps,preprint,fleqn,showpacs,showkeys]{revtex4}
\usepackage{graphicx}
\usepackage{amssymb}
\usepackage{amsmath}
\usepackage{bm}
\usepackage{amsmath, graphicx}
\usepackage{dcolumn}
\usepackage{amssymb}
\usepackage{epsfig}
\usepackage{color}
\usepackage{slashed}
\usepackage{hhline}
\usepackage{epstopdf}
\begin{document}
\newcommand{\hs}{\hspace*{0.5cm}}
\newcommand{\vs}{\vspace*{0.5cm}}
\newcommand{\be}{\begin{equation}}
\newcommand{\ee}{\end{equation}}
\newcommand{\bea}{\begin{eqnarray}}
\newcommand{\eea}{\end{eqnarray}}
\newcommand{\ben}{\begin{enumerate}}
\newcommand{\een}{\end{enumerate}}
\newcommand{\bde}{\begin{widetext}}
\newcommand{\ede}{\end{widetext}}
\newcommand{\nn}{\nonumber}
\newcommand{\crn}{\nonumber \\}
\newcommand{\Tr}{\mathrm{Tr}}
\newcommand{\non}{\nonumber}
\newcommand{\noi}{\noindent}
\newcommand{\al}{\alpha}
\newcommand{\la}{\lambda}
\newcommand{\bet}{\beta}
\newcommand{\ga}{\gamma}
\newcommand{\va}{\varphi}
\newcommand{\om}{\omega}
\newcommand{\pa}{\partial}
\newcommand{\+}{\dagger}
\newcommand{\fr}{\frac}
\newcommand{\sq}{\sqrt}
\newcommand{\bc}{\begin{center}}
\newcommand{\ec}{\end{center}}
\newcommand{\Ga}{\Gamma}
\newcommand{\de}{\delta}
\newcommand{\De}{\Delta}
\newcommand{\ep}{\epsilon}
\newcommand{\varep}{\varepsilon}
\newcommand{\ka}{\kappa}
\newcommand{\La}{\Lambda}
\newcommand{\si}{\sigma}
\newcommand{\Si}{\Sigma}
\newcommand{\ta}{\tau}
\newcommand{\up}{\upsilon}
\newcommand{\Up}{\Upsilon}
\newcommand{\ze}{\zeta}
\newcommand{\ps}{\psi}
\newcommand{\Ps}{\Psi}
\newcommand{\ph}{\phi}
\newcommand{\vph}{\varphi}
\newcommand{\Ph}{\Phi}
\newcommand{\Om}{\Omega}
\setcounter{page}{0}
\title[]{Quark Masses and Mixings in the 3-3-1 Model \\ with Neutral Leptons Based on $D_4$ Flavor Symmetry}
\author{V. V.  \surname{Vien}}
\email{wvienk16@gmail.com}
\affiliation{Department of Physics, Tay Nguyen University, 567 Le
Duan, Buon Ma Thuot, DakLak, Vietnam}
\author{H. N.\surname{Long}}
\email{hnlong@iop.vast.ac.vn}
\affiliation{Institute of Physics,
Vietnam Academy of Science and Technology, 10 Dao Tan, Ba Dinh, Hanoi, Vietnam}

\date[]{Received 6 August 2007}
\begin{abstract}
The $D_4$ flavor model based on $\mathrm{SU}(3)_C \otimes
\mathrm{SU}(3)_L \otimes \mathrm{U}(1)_X$ gauge symmetry that aims at describing quark mass and mixing is
updated. After
spontaneous breaking of flavor symmetry, with the constraint
on the Higgs vacuum expectation values (VEVs) in the Yukawa couplings, all of
quarks have consistent masses, and a realistic quark mixing matrix can be realized at the first order of perturbation theory. 
\end{abstract}



\maketitle

\section{\label{intro} INTRODUCTION}
One of the most interesting challenges in particle physics is to
determine the origin of  quark mixing, described by the unitary Cabibbo-Kobayashi-Maskawa (CKM) matrix \cite{CKM, CKM1}, which
 is approximately proportional to the identity. The CKM matrix
elements are fundamental parameters of the Standard Model (SM),
so their precise determination is important.
The CKM matrix has many parametrizations \cite{HeLiMa, Stefan, ZhangMa, FramMoh, LiMa, KangKimLee,PSWin};
however,
the CKM parametrization \cite{CKM, CKM1} and the  Wolfenstein one
\cite{Wein} are widely used.  
Recently, the discrete
symmetries have been a useful tool
 for understanding quark and lepton mixing \cite{Ahn,
HoLim, Araki, Hern, King}. The elements in the CKM matrix have now
been determined with high accuracy. The fit results for
the magnitudes of all CKM elements in Ref. 16 
imply
\begin{eqnarray}
\label{Uckml}
\left|U_{\mathrm{CKM}}\right|=\left(
\begin{array}{ccc}
0.97425\pm 0.00022          &0.2253 \pm 0.0008 &(4.13 \pm 0.49)\times 10^{-3}\\
0.225 \pm 0.008                 &0.986 \pm 0.016  &(41.1\pm 1.3)\times 10^{-3}\\
(8.4 \pm 0.6)\times 10^{-3}  &(40.0 \pm 2.7)\times 10^{-3}  &1.021\pm 0.032
\end{array}\right).
\end{eqnarray}

From Eq. (\ref{Uckml}), it follows that the quark mixing angles
are small and completely different from the lepton mixing ones that  have been studied widely by many authors in recent
years 
[17-31 and references therein].

In our previous works \cite{dlshA4, dlsvS4, dlnvS3, vlS4,vlS3,vlT7,vD4,vT7,lvZB,lvtZB}, the lepton mass
and mixing were studied in detail; however, the realistic quark mixing has not been considered.
In Ref.24, 
we studied the 3-3-1 model with neutral fermions based
    on the $D_4$ group in which the quark mixing matrix is unity at the tree-level and the $1-2$ mixing of the
ordinary quarks is obtained if the $D_4$ symmetry is violated  with $1'$; i.e, the 12 and 21 entries of the quark mixing matrix $U_{CKM}$ are non-zero if under $[\mathrm{SU}(3)_L , \mathrm{U}(1)_X, \mathrm{U}(1)_\mathcal{L}, \underline{D}_4]$ symmetries, the tensor products of fields in the quark Yukawa interactions  are $[1,0,0, \underline{1}']$ instead of $[1,0,0, \underline{1}]$ as usual. Our aim
in this paper is to construct  the 3-3-1 model with neutral
leptons based on $D_4$ flavor symmetry having a quark
mixing pattern in agreement with the most recent data.

The basic feature of the model  is that all the quark  fields act  as
different singlets under $D_4$, and
a new parametrization of quark mixing is proposed at the
tree-level. The realistic quark mixing is obtained at the
first order
 of perturbation theory when $D_4$ symmetry is violated with $1'$. 
The rest of this work is organized as follows: In Section \ref{model},
we introduce the necessary Higgs fields responsible for the charged
lepton as well as the neutrino mass and mixing. Section
\ref{quark} is devoted to quark mixing. We summarize our
results and draw conclusions in Section \ref{conclus}.
Appendix \ref{apa} presents a brief description of the $D_4$ theory.
\section{The model \label{model}}
The lepton content of the model is the same as that in Ref. 28. 
In this work, we will concentrate
 on the quark sector, where under the $[\mathrm{SU}(3)_L, \mathrm{U}(1)_X,
\mathrm{U}(1)_\mathcal{L},\underline{D}_4]$ symmetries, the left- and the right-handed quark
 fields transform as follows: \bea Q_{3L}&=& \left(u_{3L} \hs
    d_{3L} \hs
    U_{L} \right)^T\sim[3,1/3,-1/3,\underline{1}],\hs u_{3R} =[1,2/3,0,
    \underline{1}],\hs d_{3R} =[1,-1/3,0,\underline{1}],\crn
    Q_{1 L}&\equiv &
 \left( d_{1 L} \hs
  -u_{1 L}  \hs
    D_{1 L}\right)^T\sim[3^*,0,1/3,\underline{1}'], \,\,\,\, u_{1R} =[1,2/3,0,\underline{1}'],\,\,\,\,\,\, d_{1R} =[1,-1/3,0,\underline{1}'],\crn
Q_{2 L}&\equiv &
 \left( d_{2 L} \hs
  -u_{2 L}  \hs
    D_{2 L}\right)^T\sim[3^*,0,1/3,\underline{1}''],\,\,\, u_{2R} =[1,2/3,0,\underline{1}''],\,\,\,\,\, d_{2R} =[1,-1/3,0,\underline{1}''],\crn
U_R&\sim&[1,2/3,-1,\underline{1}],\hs D_{1 R}
\sim[1,-1/3,1,\underline{1}'], \,\,\,\, D_{2 R}
\sim[1,-1/3,1,\underline{1}''].\label{quarkcont}\eea
 Note that the $\underline{1}$, $\underline{1}'$ and $\underline{1}''$ for quarks meets
the requirement of the anomaly cancellation condition in the 3-3-1 models because one family of quarks
transforms differently from the others. In what follows, we consider the possibilities for generating the
quark masses. The scalar multiplets needed for this purpose
will be introduced accordingly.

To generate masses for the charged leptons, we introduce two $SU(3)_L$ scalar
triplets $\phi$ and $\phi'$, respectively, lying in $\underline{1}$ and $\underline{1}'''$ under $D_4$,
with the VEVs $\langle
\phi \rangle = (0\hs v\hs 0)^T$ and $\langle \phi' \rangle =
(0\hs v'\hs 0)^T$ \cite{dlnvS3}. From the
 Yukawa interactions
for the charged leptons, we get
$m_e=h_1v ,\, m_\mu= h v-h' v'$ and $ m_\tau=h v+h' v',$ 
  and the left- and the right-handed charged leptons mixing matrices are obtained \cite{vD4}
\bea U_l=U_R \simeq \left(%
\begin{array}{ccc}
  1 & \,\,\, 0& 0 \\
  0 &\,\,\,\frac{1}{\sqrt{2}} &\,\,\,  \frac{1}{\sqrt{2}} \\
  0 &-\frac{1}{\sqrt{2}} &\,\,\,\frac{1}{\sqrt{2}} \\
\end{array}%
\right).\label{Uclep}\eea
In similarity to the charged lepton
sector, to generate the neutrino masses, we have additionally introduced the
two scalar Higgs anti-sextets $s, \, \si$, respectively, lying in $\underline{1}$, $\underline{1}$ and $\underline{1}'$
    under $D_4$, and one $\mathrm{SU}(3)_L$ triplet lying in $\underline{1}'''$ under $D_4$, whose contribution is regarded as a small perturbation. The neutrino mass and mixing are
      then consistent with the experimental data given in Ref. 16 
in both normal and inverted hierarchical patterns. For a detailed study on the charged-lepton and neutrino sectors, the reader is referred to
 Ref. 28.

\section{Quark mass and mixing \label{quark}}
\subsection{The Tree Level \label{quarktree}}
Let's us recall the two $SU(3)_L$ Higgs scalars  responsible for charged lepton masses \cite{vD4}:
\bea \phi = \left(\phi^+_1 \, \phi^0_2 \,\phi^+_3\right)^T\sim
[3,2/3,-1/3, \underline{1}], \hs \phi' = \left(\phi'^+_1 \,
  \phi'^0_2 \,
  \phi'^+_3 \right)^T\sim [3,2/3,-1/3, \underline{1}'''].\label{phiphip}\eea
To generate the mass for quarks with a minimal Higgs content, we introduce following $SU(3)_L$ Higgs triplets: \bea
\chi&=&\left(  \chi^0_1 \hs
  \chi^-_2 \hs
  \chi^0_3 \right)^T\sim \left[3,-1/3,2/3,\underline{1}\right],\crn
  \eta&=&
\left(  \eta^0_{1} \hs
  \eta^-_{2} \hs
  \eta^0_{3} \right)^T\sim \left[3,-1/3,-1/3, \underline{1}\right],\crn
  \eta'&=&
\left( \eta'^0_{1} \hs
  \eta'^-_{2} \hs
  \eta'^0_{3} \right)^T\sim \left[3,-1/3,-1/3, \underline{1}'''\right].\label{Higgstrip}\eea
The Yukawa interactions are \bea -\mathcal{L}_q &=& h^d_3
\bar{Q}_{3L}\phi d_{3R} + h^u_1
\bar{Q}_{1L}\phi^*u_{1R} + h^u_2
\bar{Q}_{2L}\phi^*u_{2R}
+h'^u(\bar{Q}_{1L}u_{2R}+\bar{Q}_{2L}u_{1R})\phi'^*
\crn &+& h^u_3\bar{Q}_{3L}\eta u_{3R} +h^d_1
\bar{Q}_{1L}\eta^*d_{1R} +h^d_2\bar{Q}_{2L}\eta^*
d_{2R}
+h'^d(\bar{Q}_{1L}d_{2R}+\bar{Q}_{2L}d_{1R})\eta'^*\crn
&+&  f_1\bar{Q}_{1L}\chi^*
D_{1R} +f_2 \bar{Q}_{2L}\chi^* D_{2R}+ f_3 \bar{Q}_{3L}\chi U_R
+H.C.\label{Lquark}\eea

We should mention that the VEVs of $\chi,\phi,\eta$ conserve $D_4$ while those
of $\phi', \eta'$ break this symmetry into $Z_2\otimes Z_2$
\cite{vD4}. Therefore, in the quark sector, $D_4$ group is broken into $ Z_2\otimes Z_2$. We
assume that the VEVs of
$\chi, \phi, \phi', \eta,\eta'$, respectively, are given as \bea
 \langle\chi\rangle &=&
\left( 0 \hs   0 \hs   v_\chi\right)^T, \label{vevchi}\\
 \langle\phi\rangle&=&
\left( 0 \hs   v \hs   0\right)^T,\hs \langle\phi'\rangle= \left(
0 \hs   v' \hs   0\right)^T, \label{vevpp}\\
 \langle\eta\rangle&=&
\left( u \hs   0 \hs   0\right)^T, \hs \langle\eta'\rangle =
\left( u' \hs   0 \hs   0\right)^T.\label{veveep} \eea
The mass Lagrangian for quarks is then given by \bea -\mathcal{L}^{mass}_q
&=& h^d_3 v\bar{d}_{3L}d_{3R} - h^u_1 v^*
\bar{u}_{1L}u_{1R} -h^u_2 v^*\bar{u}_{2L}u_{2R}-h'^u
v'^*(\bar{u}_{1L}u_{2R}+\bar{u}_{2L}u_{1R}) \crn &+&
h^u_3 u \bar{u}_{3L}u_{3R} +h^d_1 u^*
\bar{d}_{1L}d_{1R} +h^d_2 u^*\bar{d}_{2L}d_{2R} +h'^d
u'^*(\bar{d}_{1L}d_{2R}+\bar{d}_{2L}d_{1R})\crn &+&
f_3 v_\chi \bar{U}_{L} U_R + f_1
v^*_\chi \bar{D}_{1L}D_{1R} +f_2 v^*_\chi
\bar{D}_{2L}D_{2R}+H.C.\label{Lqmass}\eea The exotic quarks
get masses \bea m_U=f_3 v_\chi ,\hs m_{D_{1,2}}=f_{1,2} v_\chi.
\eea 
The mass matrices for ordinary up- and down-quarks are,
respectively, obtained as follows: \bea M_u =
\left(%
\begin{array}{ccc}
  -h^u_1 v &\,\,\, -h'^u v' & 0  \\
   -h'^u v'  &\,\, -h^u_2v  &\, 0   \\
  \,\,\,\, 0 &\,\,\, 0 & h^u_3 u \\
\end{array}%
\right),\,\,\, M_d=
\left(%
\begin{array}{ccc}
  h^d_1 u &\,\,\,\,\, h'^d u'&\,\,\, 0  \\
   h'^d u' &\,\,\,\, h^d_2 u  &\,\,\,\,\, 0 \\
  0 & 0 & h^d_3 v \\
\end{array}%
\right).\label{MuMd} \eea The matrices $M_u, M_d$ in (\ref{MuMd})
are diagonalized as \bea V^{u+}_L M_uV^u_R=\mathrm{diag}(m_u,\,
m_c,\, m_t), \label{Mudig0}\\
V^{d+}_L M_dV^d_R=\mathrm{diag}(m_d,\, m_s,\, m_b),\label{MuMdig0}
\eea where \bea
m_u&=&-\frac{1}{2}\left[(h^u_1+h^u_2)v+\sqrt{(h^u_1-h^u_2)^2v^2+(2h'^u
v')^2}\right],\crn
m_c&=&-\frac{1}{2}\left[(h^u_1+h^u_2)v-\sqrt{(h^u_1-h^u_2)^2v^2+(2h'^u
v')^2}\right],\hs m_t=h^u_3u,\label{muct0}\\
m_d&=&\frac{1}{2}\left[(h^d_1+h^d_2)u-\sqrt{(h^d_1-h^d_2)^2u^2+(2h'^d
u')^2}\right],\crn
m_s&=&\frac{1}{2}\left[(h^d_1+h^d_2)u+\sqrt{(h^d_1-h^d_2)^2u^2+(2h'^d
u')^2}\right],\hs m_b=h^d_3v.\label{mdsb0}\eea and \bea
U^u_L&=&U^u_R=
\left(%
\begin{array}{ccc}
 \frac{K}{\sqrt{K^2+1}}&\hs-\frac{1}{\sqrt{K^2+1}}&\,0 \\
\frac{1}{\sqrt{K^2+1}}&\hs\frac{K}{\sqrt{K^2+1}}&\,0 \\
  0 &\hs 0 & \,1 \\
\end{array}%
\right), \hs U^d_L=U^d_R=
\left(%
\begin{array}{ccc}
 \frac{A}{\sqrt{A^2+1}}&\hs-\frac{1}{\sqrt{A^2+1}}&\,0 \\
\frac{1}{\sqrt{A^2+1}}&\hs\frac{A}{\sqrt{A^2+1}}&\,0 \\
  0 &\hs 0 & \,1 \\
\end{array}%
\right),\label{VuVdLR}\eea with \bea
K&=&\frac{(h^u_1-h^u_2)v+\sqrt{(h^u_1-h^u_2)^2v^2+(2h'^u
v')^2}}{2h'^u v'},\label{K}\\
A&=&\frac{(h^d_1-h^d_2)u-\sqrt{(h^d_1-h^d_2)^2u^2+(2h'^d
u')^2}}{2h'^d u'}.\label{A}\eea The CKM matrix is defined as
 \bea
U_{CKM}=U^{u}_L U^{d+}_L=\left(%
\begin{array}{ccc}
 \frac{1 + A K}{\sqrt{A^2+1}\sqrt{K^2+1}}&\frac{K-A}{\sqrt{A^2+1}\sqrt{K^2+1}}&0 \\
\frac{A - K}{\sqrt{A^2+1}\sqrt{K^2+1}}&\frac{1 + A K}{\sqrt{A^2+1}\sqrt{K^2+1}}&0 \\
 0& 0 & \,1 \\
\end{array}%
\right),\label{Vckm0} \eea where $K$ and $A$ are defined in Eqs.
(\ref{K}) and (\ref{A}).
In the special case $K=A$, i.e, \[
\frac{u}{u'}=\frac{h^u_1-h^u_2}{h^d_1-h^d_2}\frac{h'^d}{h'^u}\frac{v}{v'},\]
the $U_{CKM}$ in Eq.(\ref{Vckm0}) reduces to the identity.

In the model under consideration, the following limit is often
taken into account \cite{dlshA4, dlsvS4, dlnvS3,e331v1, e331v2}: \be u\sim u'\sim v'\sim v.
\label{limmit1}\ee
On the other hand, taking into account the discovery of the long-awaited Higgs boson at
 around 125 GeV by ATLAS \cite{Atla} and
CMS \cite{cms}, we can estimate the VEVs as follows:
\be u\sim u'\sim v'\sim v=100\,  \rm{GeV}. \label{Higsvev}\ee
The matrix $U_{CKM}$ in Eq. (\ref{Vckm0}) is closer to the
 realistic quark mixing matrix than those
derived at the tree level from other discrete symmetry
  groups \cite{dlshA4, dlsvS4, dlnvS3, vlS4,vlS3,vlT7,vD4,vT7}.
 Indeed, with the help of Eq. (\ref{Higsvev}) and by taking the
experimental data on quark mass \cite{PDG2014} \bea m_u&=&2.3\
\textrm{MeV},\hs \ m_{c}=1.275 \ \textrm{GeV},\hs m_{t}=173.21\
\textrm{GeV},\crn m_d&=&4.8\ \textrm{MeV},\hs \ m_{s}= 95
\textrm{MeV},\hs m_{b}=4.18\ \textrm{GeV},
\label{quarkmassexp}\eea as well as the average values of the CKM matrix
elements in Ref.\cite{PDG2014}
 given in Eq.(\ref{Uckml}). With
$|U_{ud}|=0.97425\pm 0.00022$, we get solutions for $A, K$
and the Yukawa quark couplings $h^u_{1,2,3}, h'^u, h^d_{1,2,3},
h'^d$ which are listed in table \ref{ModelparaII}. We see that the matrix in table
\ref{ModelparaII} is close to the realistic quark mixing matrix;
i.e, the deviations of the matrix $U_{CKM}$ in Eq.(\ref{Uckml})
from the matrix in table \ref{ModelparaII} are very small, so
this is a good approximation for the realistic quark mixing
matrix, which implies that the mixings among the quarks are
dynamically small. This is one of the most striking
predictions of the model under consideration. As we will see in section \ref{fistoder}, a violation
of $D_4$ symmetry due to Yukawa interactions will disturb
the tree-level matrix, resulting in mixing between ordinary quarks and providing the desirable quark mixing pattern.

\begin{table}[h]
\caption{ Model parameters derived from the fit with  the data
in Ref. \cite{PDG2014} at the tree - level .}
{\begin{tabular}{@{}ccccccc@{}} \toprule
  $A, K$ \hphantom{00}& $\left.%
\begin{array}{ccc}
 h^u_{1}, h^u_{2}, h^u_{3},h'^u,\\
h^d_{1}, h^d_{2}, h^d_{3}, h'^d\\
\end{array}%
\right.$ \hphantom{00}& $U_{CKM}$\\ \colrule
 $0.65558, 1.04565$ \hphantom{00}&$\left.%
\begin{array}{ccc}
-0.00610, -0.00667, 1.73500, 0.00636,\\
0.000319,0.00068, 0.04180, 0.00041\\
\end{array}%
\right.$ \hphantom{00}&$\left(%
\begin{array}{ccc}
 0.97425&0.22547&0 \\
-0.22547&0.97425&0 \\
 0& 0 & \,1 \\
\end{array}%
\right)$& \\ \colrule\botrule
\end{tabular} \label{ModelparaII}}
\end{table}
\subsection{The First-Order Corrections \label{fistoder}}
All terms of the Yukawa interactions responsible for the quarks
 masses in Eq. (\ref{Lquark}) are invariant  under the
  $[\mathrm{SU}(3)_L, \mathrm{U}(1)_X,
\mathrm{U}(1)_\mathcal{L},\underline{D}_4]$ symmetries. To obtain
a realistic quark mixing, here  we add some terms
violating $D_4$ symmetry with $1'$.  These terms are
$\widetilde{Q}_{1L}\phi^* u_{3R}, \widetilde{Q}_{1L}\eta^* d_{3R},
\widetilde{Q}_{3L}\eta u_{1R}$ and $\widetilde{Q}_{3L}\phi d_{1R}$.
Hence, the total Yukawa couplings of the ordinary
 quarks have
 two extra terms $ -{\Delta \mathcal{L}}^u_q$
 and $-{\Delta \mathcal{L}}^d_q$ which are given by
\bea -{\Delta \mathcal{L}}^u_q &=& k^u_1 \widetilde{Q}_{1L}\phi^*
u_{3R}+k^u_2\widetilde{Q}_{3L}\eta
u_{1R}+H.C,\label{deltaqu}\\
 -{\Delta \mathcal{L}}^d_q &=& k^d_1 \widetilde{Q}_{1L}\eta^* d_{3R}+
 k^d_2 \widetilde{Q}_{3L}\phi d_{1R} +H.C.\label{deltaqd}\eea
 The total mass matrices for the  ordinary up-quarks
and down-quarks then take the forms: \bea M'_u =
\left(%
\begin{array}{ccc}
  -h^u_1 v &\,\,\, -h'^u v' & -k^u_1 v \\
   -h'^u v' v &\,\, -h^u_2v  &\, 0   \\
  k^u_2 u &\,\,\, 0 & h^u_3 u \\
\end{array}%
\right),\,\,\, M'_d=
\left(%
\begin{array}{ccc}
  h^d_1 u &\,\,\,\,\, h'^d u'&k^d_1 u  \\
   h'^d u' &\,\,\,\, h^d_2 u  &\,\,\,\,\, 0 \\
  k^d_2 v & 0 & h^d_3 v \\
\end{array}%
\right).\label{Mutotal} \eea

We can separate the quark mass matrices in Eq.(\ref{Mutotal})
 into two parts as follows:
  \be M'_u = M_u + \Delta M_u,\hs M'_d = M_d + \Delta M_d,\label{Mseparate}\ee
  where $M_u$ and $M_d$ are given by
Eq. (\ref{MuMd}) due to the contributions of the invariant terms only,
and \be \Delta M_u =
\left(%
\begin{array}{ccc}
  0 &0 & -k^u_1 v \\
  0 &0  &\, 0   \\
  k^u_2 u &0& 0 \\
\end{array}%
\right),\,\,\, \Delta M_d=
\left(%
\begin{array}{ccc}
  0&0&k^d_1 u  \\
  0&0& 0 \\
  k^d_2 v & 0 & 0 \\
\end{array}%
\right)\label{DeltaMuMd} \ee are deviations from the
contributions of the $D_4$ violation terms. In the case without
$D_4$ violation, the first terms can approximately fit the data in Ref. 16
with very small deviations, as was shown in section \ref{quarktree}. The second terms belong to the contributions of
the $D_4$ violation in Eqs. (\ref{deltaqu}) and (\ref{deltaqd}).
Then, we can consider the contributions of $D_4$ violation as small
perturbations in the quark sector and terminate the theory at the
first order. At this approximation, the matrices $\Delta
M_u$ and $\Delta M_d$ in Eq. (\ref{DeltaMuMd}) do not contribute
to the quark eigenvalues. However, they change the corresponding
eigenvectors. The up- and the down-quark masses are, thus, obtained as
\bea m'_{i} = m_i\hs (i=u,c,t,d,s,b), \eea where $m_{i} \,(i=u,c,t,d,s,b)$ are given in Eqs.(\ref{muct0}) and (\ref{mdsb0}).

The unitary matrices that couple the left-handed quarks $u_L$ and
$d_L$ to those in the mass bases, respectively, are \bea U'^u_L&=&
\left(%
\begin{array}{ccc}
 \frac{K}{\sqrt{K^2+1}}&\hs-\frac{1}{\sqrt{K^2+1}}&\frac{k^u_1 v[K^2(m_c-m_t)
 +(m_u-m_t)]}{(K^2+1)(m_u-m_t)(m_c-m_t)} \\
\frac{1}{\sqrt{K^2+1}}&\hs\frac{K}{\sqrt{K^2+1}}&-\frac{K k^u_1 v(m_u-m_c)}{(K^2
+1)(m_u-m_t)(m_c-m_t)} \\
 \frac{K k^u_2 u}{\sqrt{K^2+1}(m_u-m_t)} &-\frac{k^u_2 u}{\sqrt{K^2+1}(m_c-m_t)} & \,1 \\
\end{array}%
\right), \label{UuL}\\ U'^d_L&=&
\left(%
\begin{array}{ccc}
 \frac{A}{\sqrt{A^2+1}}&\hs-\frac{1}{\sqrt{A^2+1}}&-\frac{k^d_1 u[A^2(m_s-m_b)+
 (m_d-m_b)]}{(A^2+1)(m_d-m_b)(m_s-m_b)}  \\
\frac{1}{\sqrt{A^2+1}}&\hs\frac{A}{\sqrt{A^2+1}}&\frac{A k^d_1 u (m_d-m_s)}{(A^2+
1)(m_d-m_b)(m_s-m_b)} \\
 \frac{A k^d_2 v}{\sqrt{A^2+1}(m_d-m_b)} &-\frac{k^d_2 v}{\sqrt{A^2+1}(m_s-m_b)}& \,1 \\
\end{array}%
\right),\label{UdL}\eea where $A, K$ are given in Eqs. (\ref{A})
and (\ref{K}).
 The CKM matrix at the first order of perturbation theory is now defined
 as \cite{PDG2014}
 \be
U'_{CKM}=U'^{u}_L U'^{d+}_L=\left(%
\begin{array}{ccc}
U_{11}&U_{12}&U_{13} \\
U_{21}&U_{22}&U_{23} \\
U_{31}&U_{32}&U_{33} \\
\end{array}%
\right),\label{Uckm} \ee where $U_{ij} \,\, (i,j=1,2,3)$ are given in appendix \ref{apb}.

Our model is easily shown to be consistent because the
experimental constraints on the mixing angles and the masses of
quarks can be, respectively, fitted with the quark Yukawa coupling
parameters $h^u_{1,2,3}, \, h^d_{1,2,3},\,h'^u, h'^d, k^u_{1,2},
k^d_{1,2}$ of  all the $SU(3)_L$ triplet scalars, provided that the
VEVs $u, u', v, v'$ and the quark masses are given by Eqs. (\ref{Higsvev}) and
(\ref{quarkmassexp}), respectively. Indeed, by comparing the elements of $U'_{CKM}$ in Eq. (\ref{Uckm}) with the corresponding best fit values given in Ref. 16,
we get a solution $A=-K=0.114156$, a prediction for quark mixing as presented in table \ref{parameter}, 
\begin{table}[ht]
\caption{Elements of the quark mixing matrix from the model at the first order }
{\begin{tabular}{@{}ccccccc@{}} \toprule
    Elements \hphantom{00}&The prediction \hphantom{00}& The best fit fom Ref. 16\\\colrule
$U_{ud}$ \hphantom{00} &0.94347\hphantom{00} &0.97425 \\\colrule
$U_{us}$ \hphantom{00} &0.2253\hphantom{00} & 0.2253\\\colrule
$U_{ub}$ \hphantom{00} &0.00413\hphantom{00} &0.00413 \\\colrule
$U_{cd}$ \hphantom{00} &0.2254\hphantom{00} &0.225\\\colrule
$U_{cs}$ \hphantom{00} &0.9743\hphantom{00} &0.986 \\\colrule
$U_{cb}$ \hphantom{00} &0.0411\hphantom{00} &0.0411 \\\colrule
$U_{td}$ \hphantom{00} &0.0084\hphantom{00} &0.0084 \\\colrule
$U_{ts}$ \hphantom{00} &0.04001\hphantom{00} &0.040 \\\colrule
$U_{tb}$ \hphantom{00} &0.96753\hphantom{00} &1.021 \\\botrule
\end{tabular} \label{parameter}}
\end{table}
 and \bea h^u_1&=&-1.2586\times 10^{-2},\,\,
h^u_2=-1.8672\times 10^{-4}, \,\,h^u_3=1.735,\,\, h'^u=1.43417\times 10^{-3},\label{huhd}\\
h^d_1&=&7.37265\times 10^{-4},\,\, h^d_2=2.60736\times 10^{-4},\,\, h^d_3=4.18\times 10^{-2},\,\,
h'^d=3.82925\times 10^{-4}.\nn \eea
The results in Eq. (\ref{huhd}) and table \ref{parameter} show that $h^u_1,
h^u_2, h'^u \ll h^u_3$ and $h^d_1, h^d_2, h'^d \ll  h^d_3$. There is a consequence of
 the fact that the top- and the bottom-quark masses are much larger than those of the others.

\section{Conclusion\label{conclus}}

In this paper, we have proposed a new $D_4$ flavor model based on $\mathrm{SU}(3)_C
\otimes \mathrm{SU}(3)_L \otimes \mathrm{U}(1)_X$ gauge symmetry
 in which the quark mixing matrix is concentrated. After spontaneous breaking of
 flavor symmetry, with a constraint
on Higgs VEVs in the Yukawa couplings, all of
quarks have consistent masses and a realistic quark mixing matrix can be realized at the first order of perturbation theory. Numerical estimation shows that
   the Yukawa couplings in the model under consideration are consistent with those in
   the SM.

\section*{ACKNOWLEDGMENTS}
This research is funded by the Vietnam National Foundation for
Science and Technology Development (NAFOSTED) under grant number
103.01-2014.51.

\appendix
\section{\label{apa}$\emph{D}_4$ group and Clebsch-Gordan coefficients}
$D_4$ is the symmetry group of a square. It has eight
elements divided into five conjugacy classes, with $\underline{1},
\underline{1}', \underline{1}'', \underline{1}'''$ and
$\underline{2}$ as its five irreducible representations. Any
element of $D_4$ can be formed by multiplication of the generators
$a$ (the $\pi/2$ rotation) and $b$ (the reflection) obeying the
relations $\emph{a}^4=\emph{e}$, $\emph{b}^2=\emph{e}$, and
$\emph{bab}=\emph{a}^{-1}$. $D_4$ has the following five conjugacy
classes,
\begin{eqnarray}
C_1 &:& \{a_1\equiv e\},
\crn
C_2 &:& \{a_2\equiv a^2\}, 
\crn
C_3 &:& \{a_3\equiv a,\,a_4\equiv a^3\}, 
\\
C_4 &:& \{a_5\equiv b,\,a_6\equiv a^2b\}, 
\crn
C_5 &:& \{a_7\equiv ab,\,a_8\equiv a^3b\}.\nn
\end{eqnarray}
The character table of $D_4$ is given in table \ref{D4table}, where $n$ is the order of class and $h$ is the order of elements
within each class.
\begin{table}[ht]
\caption{The character table of $D_4$}
\begin{ruledtabular}
\begin{tabular}{cccccccc}
Class & $\,n$ & $\,h$ & $\,\chi_1$ & $\,\chi_{1'}$ &
$\,\chi_{1''}$& $\,\chi_{1'''}$& $\,\chi_2$\\
\colrule
 $C_1$ & 1 & 1 & 1 & 1 & 1 & 1 &2\\
$C_2$ & 1 & 2 & 1 & 1 & 1 &1 &-2\\
$C_3$ & 2 & 4 & 1 & -1& -1& 1 &0\\
$C_4$ & 2 & 2 & 1 & 1 & -1& -1 &0\\
$C_5$ & 2 & 2 & 1 & -1& 1 &-1 &0\\
\end{tabular}
\end{ruledtabular}
\label{D4table}
\end{table}

We have worked in a real basis, in which the two-dimensional
representation $\underline{2}$ of $D_4$ is real, $2^*(1^*,
2^*)=2(1^*, 2^*)$. One possible choice of generators is given as
 \bea \underline{1}&:& a=1,\hs b=1, \crn \underline{1}' &:&
a=1,\hs b=-1,\crn \underline{1}'' &:& a=-1,\hs b=1, \crn
\underline{1}''' &:& a=-1,\hs b=-1,\crn
\underline{2} &:& a=\left(%
\begin{array}{cc}
  0 & 1 \\
  -1 & 0 \\
\end{array}%
\right),\hs b=\left(%
\begin{array}{cc}
  1 & 0 \\
  0 & -1 \\
\end{array}%
\right).\eea Using them, we calculate the Clebsch-Gordan
coefficients for all the tensor products as given below.

First, let us put $\underline{2}(1,2)$, which means some
$\underline{2}$ doublet such as $x=(x_1,x_2)\sim \underline{2}$ or
$y=(y_1,y_2)\sim \underline{2}$ and so on, and similarly for the
other representations. Moreover, the numbered multiplets such as
$(...,ij,...)$ mean $(...,x_i y_j,...)$, where $x_i$ and $y_j$ are
the multiplet components of different representations $x$ and $y$,
respectively. In the following, the components of the representations
on the left-hand sides will be omitted and should be understood, but
they always exist in order in the components of the decompositions on the
right-hand sides:
 \bea
\underline{1}(1)\otimes\underline{1}(1)&=&\underline{1}(11),\hs\underline{1}'(1)\otimes
\underline{1}'(1)=\underline{1}(11),\crn
\hs\underline{1}{''}(1)\otimes
\underline{1}{''}(1)&=&\underline{1}(11),
\hs\underline{1}{'''}(1)\otimes \underline{1}{'''}(1)=\underline{1}(11),\crn
\underline{1}(1)\otimes\underline{1}'(1)&=&\underline{1}'(11),
\hs\underline{1}(1)\otimes\underline{1}{''}(1)=\underline{1}{''}(11),\crn
\underline{1}(1)\otimes\underline{1}{'''}(1)&=&\underline{1}{'''}(11),\hs
\underline{1}'(1)\otimes\underline{1}{''}(1)=\underline{1}{'''}(11),\crn
\underline{1}{''}(1)\otimes\underline{1}{'''}(1)&=&\underline{1}{'}(11),\hs
\underline{1}{'''}(1)\otimes\underline{1}{'}(1)=\underline{1}{''}(11),\\
\underline{1}(1)\otimes\underline{2}(1,2)&=&\underline{2}(11,12),\hs
\underline{1'}(1)\otimes\underline{2}(1,2)=\underline{2}(11,-12),\crn
\underline{1}{''}(1)\otimes\underline{2}(1,2)
&=&\underline{2}(12,11),\hs
 \underline{1}{'''}(1)\otimes\underline{2}(1,2)=\underline{2}(-12,11),\\
\underline{2}(1,2) \otimes \underline{2}(1,2)
&=&\underline{1}(11+22) \oplus \underline{1}'(11-22) \oplus
\underline{1}{''}(12+21)\oplus \underline{1}{'''}(12-21).\eea 

In
the text, we usually use the following notations, for example,
$(xy)_{\underline{1}}\equiv(x_1y_1+x_2y_2)$, which is the
Clebsch-Gordan coefficients of $\underline{1}$ in the
decomposition of $\underline{2}\otimes \underline{2}$, where as
mentioned $x=(x_1,x_2)\sim \underline{2}$ and $y=(y_1, y_2)\sim
\underline{2}$. The rules to conjugate the representations \underline{1},
\underline{1}$'$, \underline{1}$''$, \underline{1}$'''$ and
\underline{2} are given by \bea
\underline{2}^*(1^*,2^*)&=&\underline{2}(1^*, 2^*),\\
\underline{1}^*(1^*)&=&\underline{1}(1^*),\,\,\,
\underline{1}'^*(1^*)=\underline{1}'(1^*),\,\,\,
\underline{1}''^*(1^*)=\underline{1}''(1^*),\,\,\,
\underline{1}'''^*(1^*)=\underline{1}'''(1^*),\nn\eea where, for
example, $\underline{2}^*(1^*,2^*)$ denotes some $\underline{2}^*$
multiplet of the form $(x^*_1,x^*_2)\sim \underline{2}^*$.

\section{\label{apb} Elements of the matrix $U'_{CKM}$}
 \bea U_{11}&=&\frac{A
K+1}{\sqrt{A^2+1}\sqrt{K^2+1}}-\frac{[K^2(m_c - m_t) + (m_u -
m_t)][A^2(m_s - m_b) +(m_d - m_b)]k^u_1
k^d_1uv}{(A^2+1)(K^2+1)(m_u-m_t)(m_c-m_t)(m_s -
m_b)(m_d-m_b)},\crn
U_{12}&=&\frac{K-A}{\sqrt{A^2+1}\sqrt{K^2+1}}+\frac{A[K^2(m_c -
m_t) + (m_u - m_t)](m_d - m_s)k^u_1
k^d_1uv}{(A^2+1)(K^2+1)(m_u-m_t)(m_c-m_t)(m_s -
m_b)(m_d-m_b)},\crn U_{13}&=&\frac{[K^2(m_c - m_t)+(m_u-m_t)]k^u_1
v}{(K^2+1)(m_u-m_t)(m_c-m_t)}+\frac{ k^d_2
v}{\sqrt{A^2+1}\sqrt{K^2+1}}\left(\frac{AK}{m_d-m_b}+\frac{1}{m_s-m_b}\right),\crn
U_{21}&=&\frac{A-K}{\sqrt{A^2+1}\sqrt{K^2+1}}+\frac{K(m_u -
m_c)[A^2(m_s-m_b)+(m_d-m_b)]k^u_1
k^d_1uv}{(A^2+1)(K^2+1)(m_u-m_t)(m_c-m_t)(m_s -
m_b)(m_d-m_b)},\crn U_{22}&=&\frac{A
K+1}{\sqrt{A^2+1}\sqrt{K^2+1}}-\frac{AK (m_u - m_c)(m_d -
m_s)k^u_1 k^d_1uv}{(A^2+1)(K^2+1)(m_u-m_t)(m_c-m_t)(m_s -
m_b)(m_d-m_b)},\crn U_{23}&=&-\frac{K k^u_1 v (m_u-m_c)}{(K^2+
1)(m_u-m_t)(m_c-m_t)}+\frac{k^d_2 v}{\sqrt{K^2+1}\sqrt{A^2+1}}\left(\frac{A}{m_d
-m_b}-\frac{K}{m_s-m_b}\right),\crn
U_{31}&=&-\frac{k^d_1 u [A^2(m_s-m_b)+(m_d-m_b)]}{(A^2+1)(m_s-m_b)(m_d
-m_b)}+\frac{k^u_2 u}{\sqrt{K^2+1}\sqrt{A^2+1}}\left(\frac{A K}{m_u-m_t}+
\frac{1}{m_c-m_t}\right),\crn
U_{32}&=&\frac{A k^d_1 u (m_d-m_s)}{(A^2+1)(m_s-m_b)(m_d-m_b)}+
\frac{k^u_2 u}{\sqrt{K^2+1}\sqrt{A^2+1}}\left(\frac{K}{m_u-m_t}-
\frac{A}{m_c-m_t}\right),\crn
U_{33}&=& 1+\frac{k^u_2k^d_2 uv}{\sqrt{K^2+1}\sqrt{A^2+1}}\left[
\frac{AK}{(m_u-m_t)(m_d-m_b)}+\frac{1}{(m_c-m_t)(m_s-m_b)}\right].\label{Uijelemen}\eea



\end{document}